\documentclass[12pt,preprint]{aastex}

\newcommand{\beq}{\begin{equation}}
\newcommand{\eeq}{\end{equation}}

\def\3he{$^3$He\,}
\def\he4{$^4$He\,}
\def\gray{$\gamma$-ray }
\def\grays{$\gamma$-rays }

\shorttitle{Derivation of Electron Distribution}
\shortauthors{Li et al.}

\begin{document}

\title{Derivation of the Electron Distribution in SNR RX J1713.7-3946 via a Spectral Inversion Method}

\author{
Hui Li\altaffilmark{1}, Siming Liu\altaffilmark{2}, and Yang Chen\altaffilmark{1,3}
}
\altaffiltext{1}{Department of Astronomy, Nanjing University, Nanjing, 210093, P.R.China}
\altaffiltext{2}{Key Laboratory of Dark Matter and Space Astronomy, Purple Mountain Observatory, Chinese Academy of Sciences,
Nanjing, 210008, P. R. China}
\altaffiltext{3}{Key Laboratory of Modern Astronomy and Astrophysics, Nanjing University, Ministry of Education, Nanjing 210093, China}

\begin{abstract}

We show that the radio, X-ray and \gray spectrum of the supernova remnant RX J1713.7-3946 can be accounted for with the simplest emission model, where all of these emissions are attributed to a population of relativistic electrons interacting with the cosmic microwave background radiation, IR interstellar photons, and a background magnetic field. With a spectral inversion method \citep{j92}, the parent electron distribution and its uncertainties are derived from the observed photon spectrum. These results are independent on the model of particle acceleration and strongly support the leptonic scenario for the TeV emission.
\end{abstract}


\section{INTRODUCTION}
\label{intro}

Relativistic cosmic rays (CRs) with energies lower than the location of the ``knee'' of the cosmic ray spectrum ($\sim10^{15}$eV) are commonly believed to be accelerated at the shock of supernova remnants (SNRs) in our Galaxy \citep{h05}. Evidence of particle acceleration by shocks of SNRs first comes from radio observations of the remnants, where the radio emission is produced by relativistic electrons via the synchrotron process. The discovery of synchrotron X-ray emission from SNR 1006 reveals acceleration of TeV electrons by these shocks \citep{k95}. However, direct evidence of proton and ion acceleration up to the spectral ``knee'' remains elusive.

Since the discovery of TeV emission from the shell-type SNR
RX J1713.7-3946, there have been debates on the nature of the dominant TeV emission mechanism \citep{e02, ah04, b06, u07, p08, b08, l08, k08, f09, z10, e10, a11, y11, i11}.
The \grays from SNRs can be produced via a hadronic process, where neutral pions produced by inelastic collisions of energetic hadrons decay into \grays. Energetic leptons can produce \grays through inverse Compton (IC) scattering of low frequency background photons. Both energetic leptons and hadrons can also produce \grays through the bremsstrahlung process \citep{b70}, which is negligible for SNR RX J1713.7-3946 \citep{a06}. Multi-wavelength observations of photon emission from this SNR in combination with theoretical considerations and/or detailed numerical modelings have been used to argue against or for the leptonic or hadronic scenarios \citep{e02, k08, p08, l08, e10, i11}. Despite all these explorations, the dominant TeV emission mechanism remains a matter of debate. Future neutrino experiments are expected to measure contributions to the observed TeV emission via the hadronic process \citep{v11}.

SNR RX J1713.7-3946 has been extensively studied from radio, IR (Acero et al. 2009), X-ray (Koyama et al. 1997; Uchiyama et al. 2003; Cassam-Chena et al. 2004), GeV \gray (Abdo et al. 2011), to TeV \gray band (Muraishi et al. 2000; Enomoto et al. 2002; Aharonian et al. 2006). The TeV \gray observations reveal a shell-like structure matching closely to the radio and non-thermal X-ray shells (Aharonian et al. 2006). Most recently, observations made with the {\it Fermi} space telescope show that the spectrum of this source in the GeV band is very hard with a power-law photon index of $\Gamma=1.5\pm0.1$, which is difficult to accommodate in a hadronic scenario, but agrees well with the IC origin of \grays in the leptonic scenario \citep{a11}. On the other hand, by considering the potentially high inhomogeneity of the shocked interstellar medium (ISM), \citet{i11} argue that a hadronic model may still explain the Fermi observation.

However, even if future observations reveal \grays via the hadronic process from isolated SNRs, this does not imply that they dominate the Galactic CR flux observed near the Earth. \citet{a09} recently showed that the de-propagated Galactic CR source spectrum is much softer than that given by most diffusive shock acceleration models. Furthermore, it has already been noticed that most of the SNRs are overlapping in superbubbles, where the particle acceleration mechanism may be quite different from that operating in isolated remnants \citep{hi98, pa04, b09, fe10}.

Given the good spatial correlation between images made at different energy bands (Acero et al. 2009), certain distributions of energetic particles or mechanisms of particle acceleration are usually introduced to fit the spatially integrated broadband spectrum in all these previous studies. Different scenarios of the background radiation field and ISM are also considered. Uncertainties in the particle acceleration process at the SNR shocks, especially in the spatial diffusion coefficient of energetic particles and injection mechanisms of supra-thermal particles, have made the relevant studies inconclusive \citep{f10}.

In this Letter, we show that the current observations, though very extensive, may not justify these sophisticated modelings.
Assuming that the TeV \grays originate from the IC by relativistic electrons of the cosmic microwave background radiation (CMB) and an IR background with a black-body spectrum suitable for the Galactic environment of the remnant, we show that one can derive the parent electron distribution and its uncertainties directly from \gray observations with a well-established spectral inversion method (\S\ \ref{method}). It is also shown that by adjusting the magnetic field and the extrapolation of the parent electron distribution toward low- and high-energies, the radio to X-ray spectrum can also be reproduced. Therefore the simplest emission model, where the broadband emission of SNR RX J1713.7-3946 is attributed to a population of relativistic electrons interacting with a background magnetic field, the CMB and an IR photon background, can fully account for the radiation spectrum and gives direct constraints on the particle acceleration processes (\S\ \ref{results}). Although these results do not preclude some level of proton \grays being masked by the more intense electron \grays, improved spectral measurements are needed to go beyond such a simple model. Conclusions and implications of our study are drawn in \S\ \ref{conclusion}.

\section{SPECTRAL INVERSION METHOD}
\label{method}

The primary goal of the study of astrophysical sources is to use characteristics of the observed emissions to probe the underlying physical processes. In general, the related radiation mechanisms are the first and most important processes to be studied. When the quality of the observational data is not sufficiently good, some forward modeling approach is usually taken to explain the relevant observations. The consequent results are usually model dependent since different assumptions can be introduced in the proposed models.

When the relevant radiative mechanisms are well established and the observed data have a high quality, a spectral inversion method may be used to derive the distribution of the parent particle distribution from the observed radiation spectrum. The parent particle distribution is more directly connected to the underlying physical processes than the observed radiation spectrum. This is a classical inversion problem and has been studied extensively in solar physics \citep{j92}. In the hadronic scenario for the TeV emission from shell-type SNRs, \citet{v06} shows that the observed TeV emission spectrum can be used to derive the neutrino spectrum from these sources. However, it is well known that the leptonic scenario for the TeV emission from SNRs has fewer model parameters and faces less challenges than the hadronic one \citep{l08, k08, e10}. In light of recent simplifications of the treatment of the IC process \citep{p09}, the spectral inversion method can be readily applied to the leptonic scenario for the TeV emission from a few shell type SNRs. In the following, we will only consider the CMB and an IR background as seed photons for the IC process.

\subsection{IC Radiation of Mono-energetic Electrons in a Black-body Photon Field}\label{sec:IC-approx}

\citet{p09} has explored the IC radiation of mono-energetic electrons in  black-body photon fields. For the sake of completeness, we summarize the key results relevant to our study in the following.
The spectral distribution of the volume emissivity of an isotropically distributed
electron population due to the IC process is given by
(Blumenthal \& Gould 1970)
\begin{equation}\label{PICdef}
P(k)=ck\int d\gamma N(\gamma)\int d\epsilon n_{\rm ph}(\epsilon)\sigma_{\rm IC}(k,\epsilon;\gamma)
\end{equation}
where $\epsilon$ and $k$ are the photon energies before and after the interaction, respectively, $c$, $\gamma$, $N(\gamma)$, $n_{\rm ph}(\epsilon)$, and $\sigma_{\rm IC}$ are the speed of light, the Lorentz factor, spectral distribution of electrons, the energy distribution of the background photons, and the angle-integrated IC cross-section, respectively.
Equation~(\ref{PICdef}) can be rewritten as
\begin{equation}\label{ICsim}
P(k)=\int d\gamma N(\gamma)p(\gamma,k),
\end{equation}
where $p(\gamma,k)$ represents the spectral distribution of IC radiation of mono-energetic electrons with a Lorenz factor $\gamma$.

In some astrophysical cases, $n_{\rm ph}(\epsilon)$ can be represented by the isotropic blackbody radiation
$
n_{\rm ph}(\epsilon)={8\pi\epsilon^2}/h^3c^3[{\rm exp(\epsilon/\epsilon_{\rm c})-1}]
$
where $\epsilon_{\rm c}=k_{\rm B}T$, and $k_{\rm B}$, $h$, and $T$ are the Boltzmann constant, the Planck constant, and the temperature of the radiation field, respectively.
Then $p(\gamma,k)$ in Equation~(\ref{ICsim}) can be approximated as
\begin{equation}\label{IC-single}
p(\gamma,k)=\frac{6\pi\sigma_{\rm T}m_{\rm e}^2c^2\epsilon_{\rm c}}{h^3\gamma^{2}}I(\eta_{\rm c},\eta_0)
\end{equation}
with the function $I(\eta_{\rm c},\eta_0)$ given by
\begin{equation}\label{I-approx}
I\approx\frac{\pi^2}{6}\eta_{\rm c}\rm exp\bigg[-\frac{2\eta_0}{3\eta_{\rm c}}-\frac{5}{4}\left(\frac{\eta_0}{\eta_{\rm c}}\right)^{1/2}\bigg]
+\frac{\pi^2}{3}\eta_{\rm c}\eta_0\rm exp\bigg[-\frac{2\eta_0}{3\eta_{\rm c}}-\frac{5}{7}\left(\frac{\eta_0}{\eta_{\rm c}}\right)^{0.7}\bigg],
\end{equation}
where $\eta_{\rm c}\equiv\epsilon_{\rm c}k/(m_{\rm e}c^2)^2$ and $\eta_0\equiv k^2/[4\gamma m_{\rm e}c^2(\gamma m_{\rm e}c^2-k)]$, $m_e$ is the electron rest mass (Petruk 2009).
With the approximation above, the double integral in Equation~(\ref{PICdef}) for calculating the IC emissivity is simplified into a single integral as given by Equation~(\ref{ICsim}). This simplification facilitates further steps in our spectral inversion method significantly. When there are multi-blackbody components, one can obtain an $I$ for each temperature and add them together with the appropriate weight of their photon energy density to get the overall $p$.

\subsection{Matrix Formulation of the IC Emission}
\label{sec:IC-mat}

To derive the electron distribution $N(\gamma)$ from the observed photon spectrum, which is proportional to $P(k)$, one needs to use energy bins of the observed photon spectrum to discretize Equation (\ref{ICsim}) (Johns \& Lin 1992):
\begin{equation}\label{eq:discretize}
P(k_i)=\sum^n_{j=i} \bigg(\int^{\gamma_{j+1}}_{\gamma_{ j}}N(\gamma)p(\gamma,k_i)d\gamma\bigg)
=\sum^n_{j=i} \bigg(\overline{N}(\gamma_{j}) \int^{\gamma_{j+1}}_{\gamma_{ j}}p(\gamma,k_i)d\gamma\bigg)
\end{equation}
where $k_n$ corresponds to the highest energy of the observed photons, $\gamma_{i}$ is the minimum Lorentz factor of electrons that can contribute to the IC \grays with an energy $k_i$ \footnote{In practice, at a given Lorentz factor $\gamma$, $p(k)$ may peak at several photon energies if there are multi-blackbody components. We adjust $\gamma$ so that the highest photon energy, where $p(k)$ peaks, is equal to $k_i$. Then $\gamma_{i}$ is chosen as a factor of 1.3 lower than this Lorentz factor.}.
$\overline{N}(\gamma_{j})$, which is what we will obtain directly from the inversion method, is the electron density in the Lorentz factor interval ($\gamma_{j}$, $\gamma_{j+1}$) averaged over $p(\gamma,k_i)$.

So, for the $n$ energy bins of the observed photons, $P(k_i)$ forms an $n\times1$ matrix, and the de-convolved electron distribution, $\overline{N}(\gamma_{j})$, forms another $n\times1$ matrix. Then Equation (\ref{eq:discretize}) can be expressed in a matrix form:
\begin{equation}
P({k_i})=\sum_{j=i}^n\beta_{ij} \overline{N}({\gamma_{j}})\,,
\end{equation}
where the IC power matrix $\beta_{ij}$ is given by:
\begin{equation}\label{beta-M}
\beta_{ij}=\left\{
\begin{array}{ll}
\int_{\gamma_{j}}^{\gamma_{j+1}} p(\gamma,{k_{i}}){\rm d}\gamma
=\int_{\gamma_{j}}^{\gamma_{j+1}}{\rm d}\gamma\ 6\pi\sigma_{T}m_{\rm e}^2c^2\epsilon_{\rm c}h^{-3}\gamma^{-2}I(\eta^{i}_{\rm c},\eta^{i}_0) \quad &{\rm for} \quad j\ge i\,, \\
0 \quad &{\rm for} \quad j < i\,.
\end{array}
\right.
\end{equation}
Then the electron distribution can be derived by inverting this matrix equation
\begin{equation}\label{eq:eletron-Inv}
\overline{N}(\gamma_{i})=\sum_{j=i}^n \alpha_{ij} P(k_{j}),
\end{equation}
where the matrix $\alpha_{ij}$ is the inversion of the matrix $\beta_{ij}$:
\begin{equation}\label{alpha-M}
\alpha_{ij}=\left\{
\begin{array}{ll}
\sum_{l=i+1}^j-{\beta_{ii}^{-1}\beta_{il} }\alpha_{lj} \quad &{\rm for} \quad j> i\,, \\
\beta_{ii}^{-1}\quad &{\rm for} \quad j=i\,, \\
0 \quad &{\rm for} \quad j < i\,.
\end{array}
\right.
\end{equation}
This kind of inversion method was first introduced by \citet{j92} to derive the parent electron distribution from the optically thin spectrum of bremsstrahlung photons. The IC of blackbody spectra has similar characteristics as the bremsstrahlung process after adopting the approximation discussed in \S~\ref{sec:IC-approx}. One therefore can readily adopt the formulas in this method.
Note that $\alpha_{ij}$ is given explicitly here, and there should be a minus sign in the original Equation (14) given by \citet{j92}.

Since there are uncertainties in the observed \gray\ spectrum, the propagation of errors from the photon spectrum to the derived electron spectrum should be considered carefully. Fortunately, the matrix Equation (\ref{eq:eletron-Inv})
offers a very straightforward way of calculating the errors in the electron spectrum. The densities of electrons simply linearly depend on the photon fluxes, the errors of the electron densities are then given by:
\begin{equation}\label{eq:error}
\delta \overline{N}(\gamma_{i})=\bigg[\sum^n_{j=i}\bigg(\alpha_{ij}\delta P_{\gamma}(k_j)\bigg)^2\bigg]^{1/2}.
\end{equation}
Thus, we can use this matrix formulation to obtain the electron distribution and errors from the observed photon fluxes and errors.

\section{RESULTS}
\label{results}
We now apply our inversion method to the GeV and TeV observations of SNR RX J1713-3946 by the Fermi-LAT and HESS \citep{a11, a06}, respectively.
Besides the CMB, we assume IR photons with a temperature of 30 K and an energy density of 1 eV cm$^{-3}$ following \citet{p06}.
Figure \ref{fig:electron} shows the \gray spectrum of this SNR along with the resulting electron spectrum after applying the inversion method. Due to large errors and fluctuations of the observed \gray fluxes, negative electron densities are inferred for some energy intervals, which is clearly not physical but mathematically expected. To address this issue, a running smooth of the observed fluxes is done before applying the inversion method and the corresponding data are indicated by the squares in the Figure. Even with such a smoothing, the error of the derived electron density is still large in some energy bins. Future observations with improved \gray flux density data will give a better electron spectral measurement.

To verify the validity of this method, one may calculate the IC and synchrotron spectra of the obtained electron distribution and compare them with observations. To do this properly, one first needs to interpolate the obtained electron spectrum since the energy bins are wide and the radiation spectra are sensitive to spectral details. We use a 2nd-order polynomial to do the interpolation. To calculate the synchrotron spectrum, one also needs to specify the magnetic field and do the extrapolation of the electron distribution toward low and high energies.
We assume power-law distributions for these extrapolations with the power-law indices as free parameters. By adjusting the magnetic field and the two power-law indices, one can fit the observed radio to TeV spectrum,
as shown in Figure~\ref{fig:multiband}. For the  best fit, the value of the magnetic field $B=15\ \mu$G and the values of the low- and high-energy end electron spectral indices are 2.1 and 10, respectively.
This smooth electron distribution is shown as the red solid line in the right panel of Figure~\ref{fig:electron}.

\section{DISCUSSIONS AND CONCLUSIONS}
\label{conclusion}

\begin{figure}[ht]
\begin{center}
\includegraphics[width=8cm]{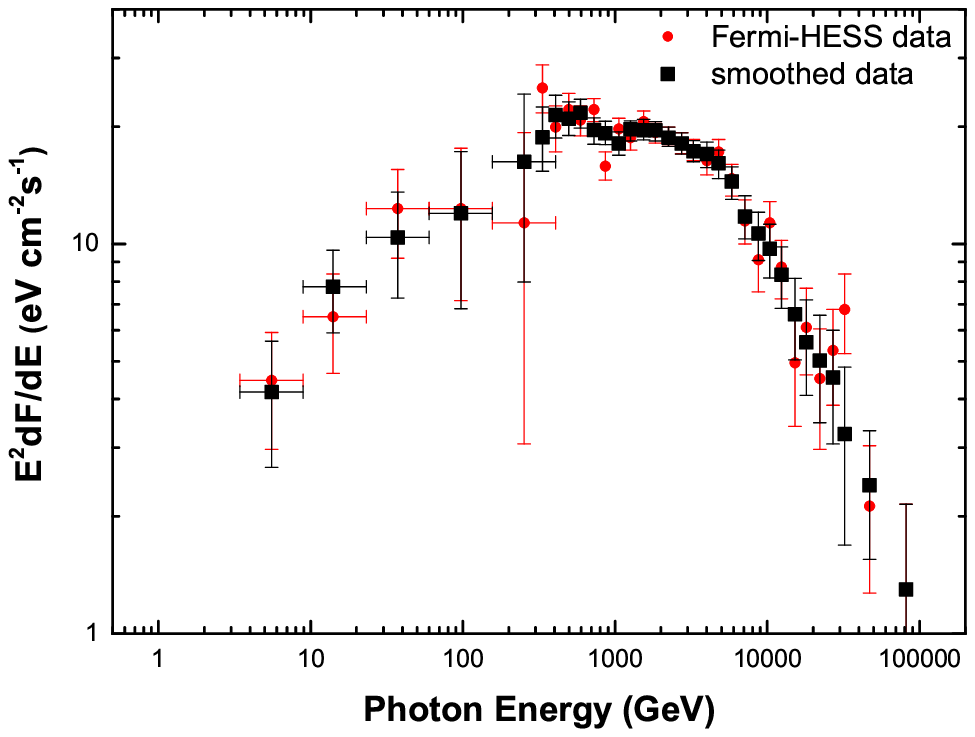}
\includegraphics[width=8cm]{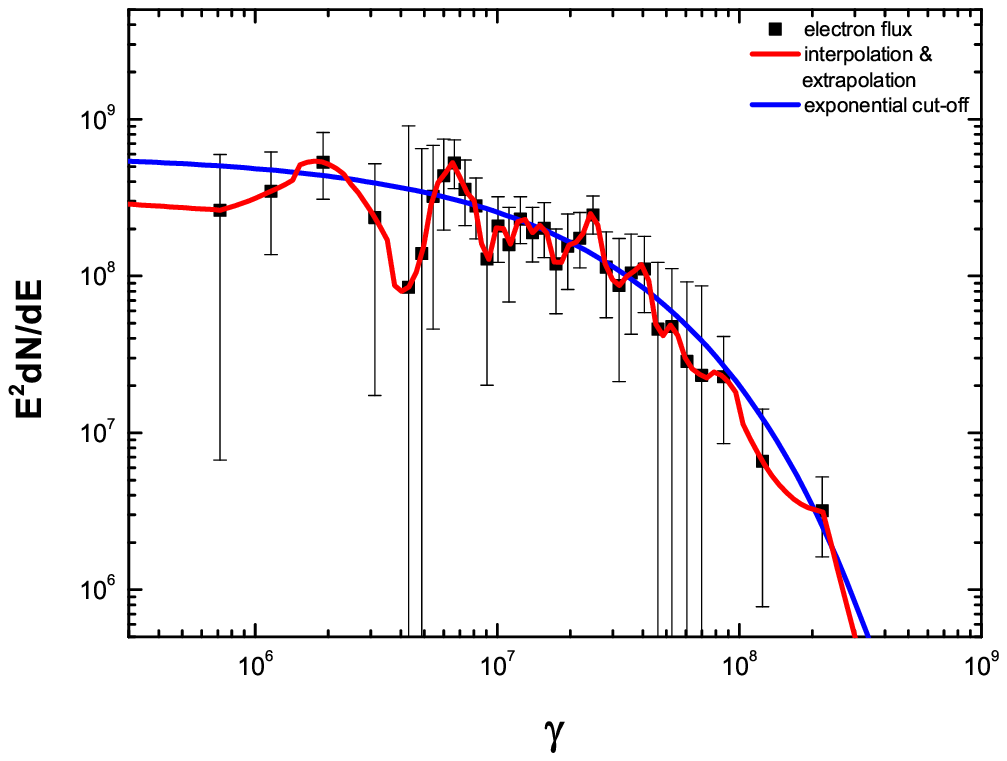}
\caption{Left: the observed (circle) and smoothed (square) GeV and TeV \gray fluxes for RX J1713.7-3946. (Abdo et al. 2011, Aharonian et al. 2006). Right: the derived electron distribution with error bars. The red solid line represents the inter- and extrapolated electron distribution, which is used to calculate the synchrotron and IC radiation spectra in Figure \ref{fig:multiband}. The blue solid line shows an analytical function, which can also reproduce the observed radiation spectrum.}
\label{fig:electron}
\end{center}
\end{figure}

\begin{figure}[ht]
\begin{center}
\includegraphics[width=15cm]{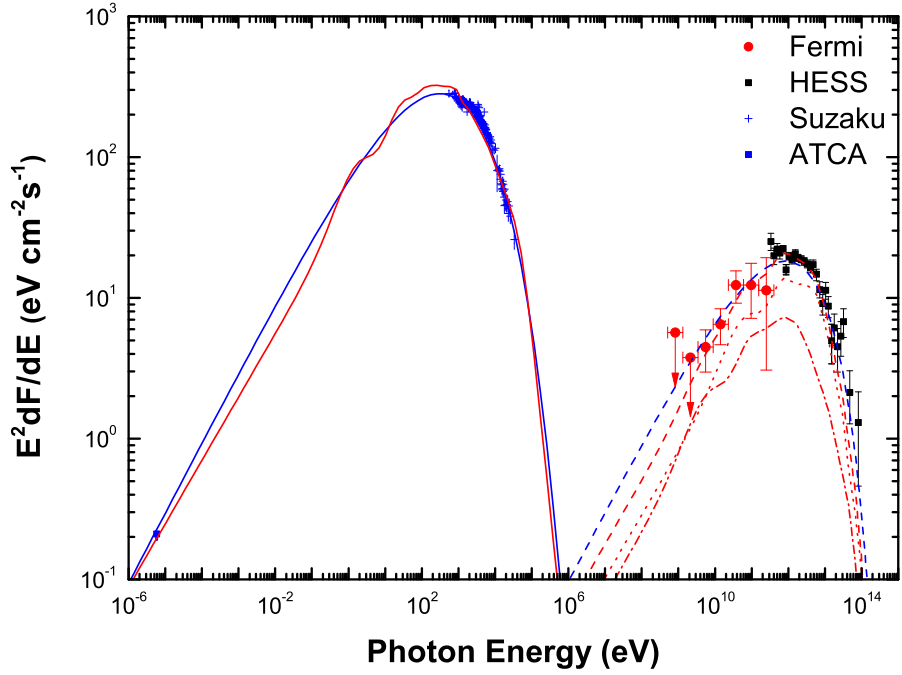}
\caption{
Comparison of the observed radio (Acero et al. 2009), X-ray (Tanaka et al. 2008) and \gray fluxes with the synchrotron (solid) and IC (dashed) spectra of the derived electron distributions using our inversion method. The blue lines are for the analytical distribution, whose parameters are described in \S\ ~\ref{results}. The red lines are for the inter- and extrapolated electron distribution, where the dotted and dot-dashed lines are for the IC of IR and CMB photons, respectively. }
\label{fig:multiband}
\end{center}
\end{figure}

With a spectral inversion method, we show full consistency of the simplest leptonic model for the radio to TeV spectrum of SNR RX J1713.7-3946. The model does not prescribe the particle acceleration process and assumes only the CMB and an IR background as the seed photons for the IC scatter. 
The current observations, though extensive, therefore may not justify very sophisticated modeling of this source, and there is no prominent evidence of emission from energetic hadrons.
The errors of the derived electron distribution are large and the resulting electron distribution can be readily fitted with a simple analytical function: $\gamma^{-2}\exp{[-(\gamma/1.3\times 10^7)^{0.6}]}$ (blue solid line in the right panel of Fig. \ref{fig:electron}), which also fits the overall radiation spectrum (blue lines in Fig. \ref{fig:multiband}). Although the difference between this analytical function and the distribution obtained from the inter- and extrapolation of the derived electron densities appears to be prominent, it is not statistically significant due these large errors.

These results have not taken into account the effect of optical background photons. Inclusion of this component, though more involving, is straightforward. Actually, even the adoption of blackbody spectra is not necessary. With the blackbody background photon spectrum, one may obtain some analytical expressions to better appreciate these results. For the spectral inversion method to be applicable, one just requires that, for a given Lorentz factor of electrons, the IC spectrum cutoff sharply at high energies so that the IC cross section may be represented by a triangle matrix. To estimate the effect of the optical background photons, we derive the electron distribution assuming the CMB alone. With a lower overall photon energy density, the inferred magnetic field is only 8.5$\mu$G, which is expected. The analytical approximation of the electron distribution is very similar except that the cutoff Lorentz factor changes from $1.3\times 10^7$ to $1.6\times 10^7$, which is caused by the higher energy of IR photons than that of the CMB. The spectral indexes of the electron distribution extrapolated to low and high energies are 2.2 and 10, respectively, which are essentially the same. Therefore, with an extra optical photon background, we expect that the magnetic field increases even more and the cutoff energy of the electron distribution decreases slightly.

The analytical function is similar to those used in previous more quantitative studies \citep{f10, y11}. However, a $\chi^2$ evaluation of the goodness of the model may not be justified for the following reasons: 1) there are systematic errors in the observed X-ray and \gray fluxes; 2) the Suzaku X-ray spectrum is obtained from a part of the remnant and re-scaled to give the overall fluxes \citep{2008ApJ...685..988T}, which may not be well justified due to the asymmetry of the source \citep{2009A&A...505..157A}; 3) the complex source structure also betrays the simple one zone emission model \citep{2009A&A...505..157A}, and both the magnetic field and the electron distribution may vary significantly across the remnant \footnote{Although it is well established that there are filaments of strong magnetic field in SNRs \citep{u07}, the filling factor of these filaments is usually low and their contributions to the spatially integrated spectrum are consequently low. Our simple one zone model for the bulk spectrum of the source is therefore justified. However, for studies directly related to these filaments, the inhomogeneity of the magnetic field must be considered \citep{bk09}.}; 4) the background radiation field is not well known, which constitutes another systematic uncertainty; 5) the particle acceleration process can be intrinsically complex without simple distributions of the acceleration particles \citep{p04, m01, c08, b09, fe10, ad11}. Nevertheless, the method can be used to other sources where the emission is dominated by optically thin IC process.






\acknowledgements
We thank Dr. Vissani, F. and the anonymous referee for helpful comments and discussions. This work is supported by the National Natural Sciences Foundation of China via the grants 11143007, 11173064, and 10725312 and the 973 Program via the grant 2009CB824800.

{}

\clearpage

\clearpage

\end{document}